\documentclass{epl}

\title{Slow Light amplification in a non-inverted gain medium}
\author{F. Caruso, I. Herrera$^{1}$, S. Bartalini$^{1}$ and F.~S.~Cataliotti$^{1,2}$}
\institute{
  Laboratorio di Informazione Quantistica, Scuola
Superiore di Catania, via S. Paolo 73, I-95123 Catania, Italy.\\
\inst{1} European Laboratory for Non-Linear Spectroscopy (LENS),
via Nello Carrara 1, I-50019 Sesto Fiorentino (FI), Italy.\\
  \inst{2} Dipartimento di Fisica, Universit\`a di Catania, via S.
Sofia~64, I-95124 Catania, Italy.\\
  }


\begin{document}

\maketitle

\begin{abstract}
We investigate the propagation of a coherent probe light pulse
through a three-level atomic medium (in the
$\Lambda$--configuration) in the presence of a pump laser under
the conditions for gain without inversion. When the carrier
frequency of the probe pulse and the pump laser are in a Raman
configuration, we show that it is possible to amplify a slow
propagating pulse. We also analyze the regime in which the probe
pulse is slightly detuned from resonance where we observe
anomalous light propagation.
\end{abstract}

The intriguing physics of atomic three--level systems, for the first time observed in Pisa in 1976
by the group of A. Gozzini \cite{alzetta}, is at the basis of all the recent experiments on slow
light propagation \cite{hau} and speculations about possible realizations of quantum memories
\cite{fleischhauer}, quantum phase gates \cite{vitali} and photon--counters with unprecedented
efficiency \cite{imamoglu,lukinRMP}. Many other phenomena involving three-level systems have
attracted much interest; among them the possibility of observing amplification without inversion
(AWI) \cite{ari} and anomalous propagation \cite{artoni,godone}. Anomalous propagation but in an
inverted two--level system has already been studied by the group of R. Chiao \cite{bolda} and more
recently in a three--level configuration with Raman scheme by the group of Dogariu \cite{dogariu}.
On the other hand subluminal propagation inside the gain band of a xenon discharge has been
observed early in the seventies \cite{casperson}.

In this letter we extend the treatment of \cite{artoni} to the
case of a pumped $\Lambda$--configuration under the conditions for
gain without inversion. This offers the possibility to observe
gain while at the same time reducing the propagation speed of a
pulse. Furthermore, it is possible to reduce absorption in the
anomalous propagation regime. Indeed, in the latter regime, even
though the pulse center of mass is advanced with respect to a
pulse propagating in vacuum, the pulse edge never goes faster than
the speed of light in vacuum, unless population inversion is
present. In order to observe gain without inversion it is
necessary to populate both the ground states of the system without
populating the excited state. We show how this can be achieved
starting from the thermal population (equal population in the
ground states and no population in the excited state) by
introducing suitable incoherent pumping rates in the ground
states.

\begin{figure}[t!]
\onefigure[width=.4\textwidth ]{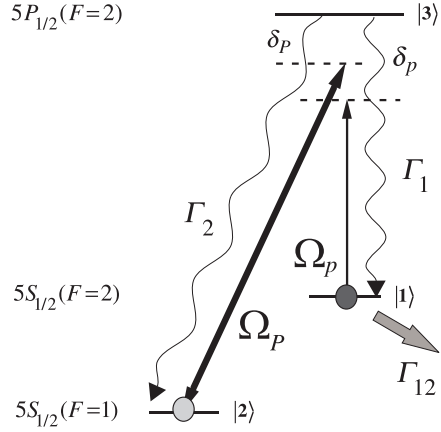} \caption{Level scheme for the $^{87}Rb$ $D_{1}$
line. The three levels are coupled by a pump laser with a Rabi frequency $\Omega_{P}$ and a probe
field with a Rabi frequency $\Omega_{p}$. Only the hyperfine ground sub-levels $|{1} \rangle$ and
$|{2} \rangle$ can be initially populated. The relevant linewidths are $\Gamma_{1}/2\pi \simeq
\Gamma_{2}/2\pi \simeq \gamma_{3}/2 \pi= 5.75 $ MHz while $\Gamma_{12}/2\pi \simeq 1$ KHz.}
\label{fig:scheme}
\end{figure}

We start by considering a pulse, Gaussian in form, propagating through an atomic medium of optical
thickness $d$ which is much smaller than the incident pulse length $\cal {L}$ \cite{artoni}. The
level scheme we analyze is represented in figure \ref{fig:scheme} and consists of two ground
sub--levels $|{1}\rangle$ and $|{2}\rangle$ and an excited level $|{3}\rangle$. A strong laser
field (pump) with Rabi frequency $\Omega_P$ and a weak laser field (probe) with Rabi frequency
$\Omega_p$ connect to the excited level the ground sub--levels $|{2}\rangle$ and $|{1}\rangle$
respectively. The excited state $|{3}\rangle$ can decay to the ground states $|{1}\rangle$ and
$|{2}\rangle$ with rates $\Gamma_1$ and $\Gamma_2$ respectively. Only ground state $|{1}\rangle$
decays to levels outside those considered here with rate $\Gamma_{12}$.

As in \cite{artoni} we analyze propagation using the lowest order
expansion in the probe intensity of the complex refractive index
$n(\omega)$ around the carrier frequency $\omega_c$. This
approximation is fully justified in the regime considered, as it
will be evident in the following discussion. We are interested in
the group velocity $v_g$, its dispersion $d_g$ and the probe
transmission intensity $G_T$. In the above expansion these
quantities are represented by:
 \begin{equation}
\label{eq:vg} v_g=\frac{c}{\eta(\omega)+\omega
\frac{\partial\eta(\omega)}{\partial\omega}}
 \end{equation}
 \begin{equation}
\label{eq:dg} d_{g}(\omega)= -\frac{v_{g}^{2}(\omega)}{c} \left(\omega \
\frac{\partial^2\eta(\omega)}{\partial\omega^2}+2
\frac{\partial\eta(\omega)}{\partial\omega}\right)=- v_{g}^{2}(\omega)\frac{
\cal{D}(\omega)}{\Gamma_1}
\end{equation}
 \begin{equation}
 G_{T}(\omega)=|T(\omega)|^{2}=\Bigg|\frac{4
n(\omega)}{\left[n(\omega)+1\right]^{2}-\left[n(\omega)-1\right]^{2}
e^{ 2i n(\omega)\omega d/c}}e^{i \left[n(\omega)-1\right]\omega
d/c}\Bigg|^{ \ 2} \label{eq:T}
 \end{equation}
where $\eta(\omega)$ is the real part of the refractive index $n(\omega)$, $d$ is the length of the
atomic medium and $c$ the speed of light in vacuum. We have defined a group velocity dispersion
function $\cal{D}(\omega)$ that has the dimension of a reciprocal velocity.

We use a density matrix treatment and expand in  powers of the
probe Rabi frequency. For weak probe intensities the complex
steady-state atomic susceptibility exhibited to the probe can be
fully accounted for by the first order expansion,
\begin{eqnarray}
\label{eq:chi}
\chi_{p}& = &  3 \pi  {\cal{N}}_{p} \Gamma_{1}
\Big[\frac{ (\delta_{p} -i \gamma_{1})} {(\delta_{p} - i
\gamma_{3})(\delta_{p} -i
\gamma_{1})-(\Omega_{P}/{2})^{2}}(n_{1}-n_{3})-{}
\nonumber\\
& & {}-\frac{i}{\gamma_{3}} (\Omega_{P}/{2})^{2}
\frac{(n_{3}-n_{2})} {(\delta_{p} - i \gamma_{3})(\delta_{p} -i
\gamma_{1})-(\Omega_{P}/{2})^{2}} \Big]
 \end{eqnarray}
where ${\cal{N}}_{p}$ is the scaled sample average density $(\lambda_{p}/2 \pi)^{3} N /V$,
$\lambda_{p}$ is the probe resonant wavelength and $n_i$ is the normalized population of level $i$.
The overall dephasings $\gamma_{3}=(\Gamma_{1}+\Gamma_{2})/2$ and $\gamma_{1}=\Gamma_{12}/2$ of
levels $|{3}\rangle$ and $|{1}\rangle$ are expressed in terms of the respective levels linewidths
(see fig. \ref{fig:scheme}). In equation (\ref{eq:chi}) we also neglect contributions due to the
atomic velocity distribution, since these can be eliminated by choosing a co--propagating pump and
probe laser configuration.

\begin{figure}[!]
\onefigure[width=.5\textwidth ]{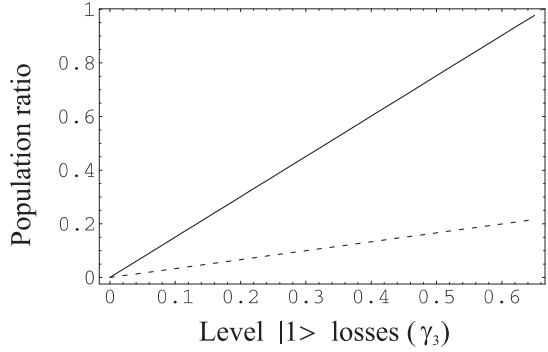} \caption{Population ratio between levels
$|{2}\rangle$ and $|{1}\rangle$ (continuous line) and $|{3}\rangle$ and $|{1}\rangle$ (dashed line)
as a function of the losses from level $|{1}\rangle$.} \label{fig:populations}
\end{figure}

In obtaining equation (\ref{eq:chi}) we have assumed that the populations of the levels do not vary
much with time. This assumption is justified since the Rabi frequency of the pump laser
$\Omega_{P}$ is smaller than the excited state linewidth $\gamma_3$ ($\Omega_{P}< \gamma_{3}$), and
the probe laser is taken to be much weaker than the pump. Furthermore, because  the excited state
$|{3}\rangle$ may decay to other atomic levels outside the two considered here, in all the
following we will assume $n_{3}=0$. In order to check the validity of these assumption we have
solved the full system of optical Bloch equations for an open three level system including the
possibility to pump population in any of the three levels and the possibility for the population to
decay to levels outside those considered here from any of the three levels. The usual situation
corresponds to the case where decay from the ground levels is negligible with respect to decay from
the excited state. However, in such a case the steady state population ratio of the two ground
levels is fixed by the Rabi frequencies of the lasers. Instead we consider the more realistic
situation of a room temperature cell in which atomic populations are thermally distributed among
all levels. Therefore the initial configuration would be a nearly 50/50 distribution between the
two ground states; in addition we have introduced a loss mechanism from the level $|{1}\rangle$
while for level $|{3}\rangle$ we have considered the correct branching ratios for $^{87}$Rb. In
figure \ref{fig:populations} we show how the steady state population ratio between the two ground
levels, even in presence of the laser beams, can indeed be varied this way while, at the same time,
the amount of population placed in the excited level remains negligible. We note that a similar
loss from the ground state can be easily implemented by an incoherent RF field stimulating a
transition to any other ground sub--level. However this configuration has the disadvantage that
while the population in level $|{2} \rangle$ increases so does the dephasing rate $\gamma_1$ of the
ground states superposition. This strongly affects both the gain and the propagation of the pulse
as discussed below.

\begin{figure}[!]
\includegraphics[width=.47\textwidth ]{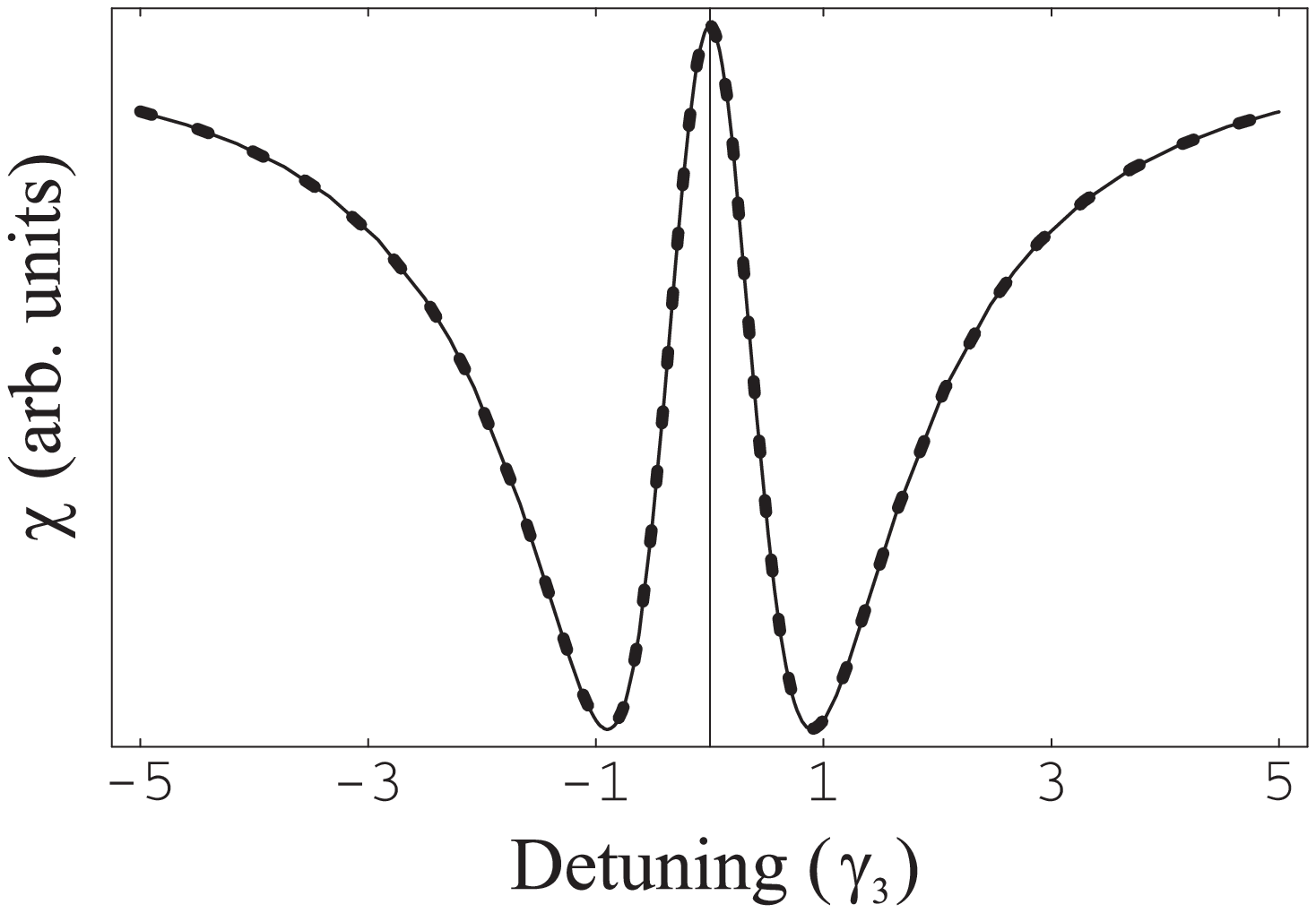}
\includegraphics[width=.52\textwidth ]{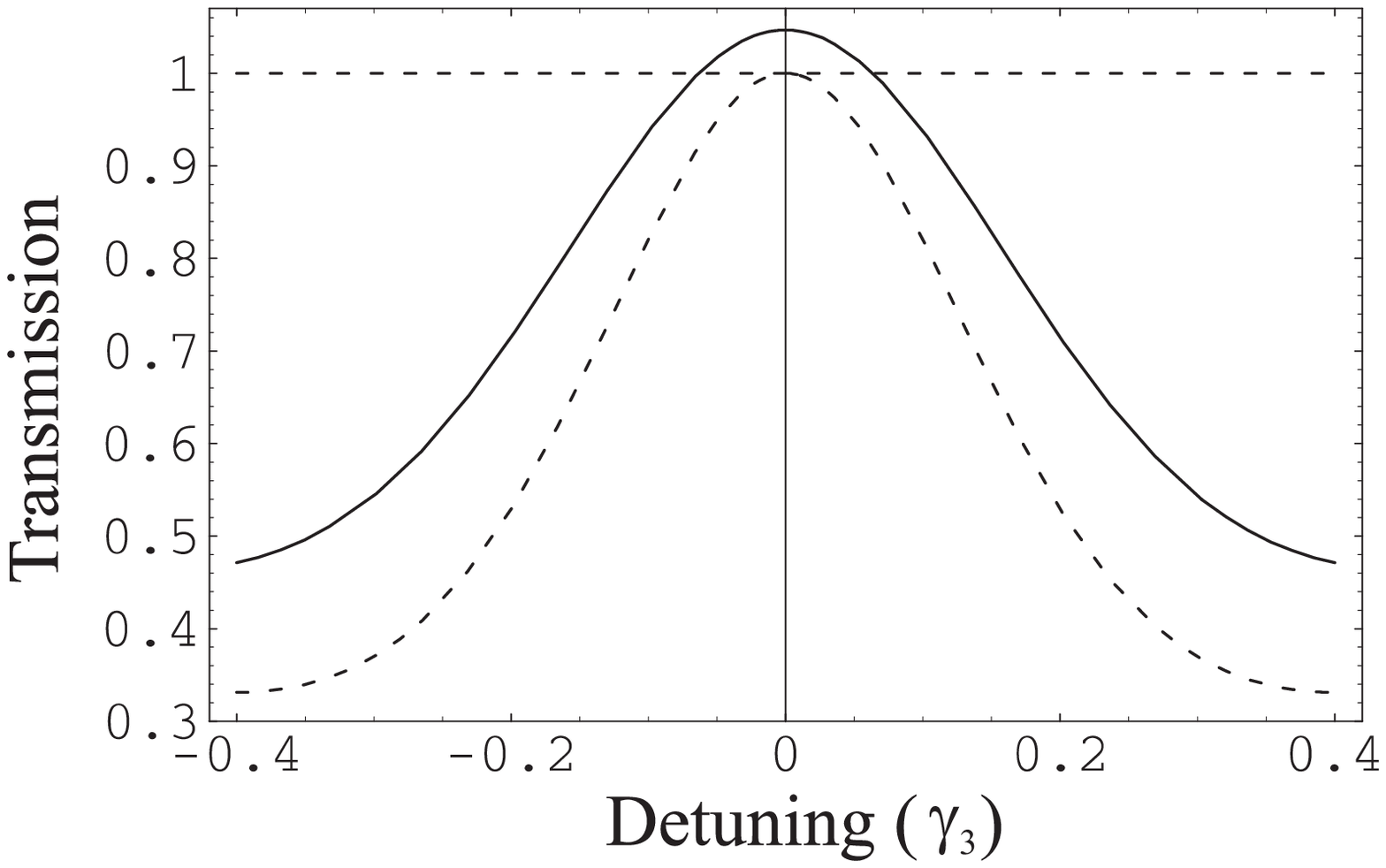}
\caption{Left: Imaginary part of the atomic susceptibility from the full density matrix treatment
for a loss rate from level $|{1}\rangle$ of $0.2 \gamma_3$ as a function of the probe detuning
$\delta_p$ (continuous line). Atomic susceptibility from equation (\ref{eq:chi}) when $30\%$ of the
population is placed in level $|{2}\rangle$ as a function of probe detuning $\delta_p$ (dashed
line). In both cases the pump Rabi frequency was $0.8 \gamma_3$. Right: Probe transmission as a
function of probe detuning for a 10 cm long cell containing rubidium at $35^\circ$ C, when all
population is in level $|{1} \rangle$ (dashed line) and when $30\%$ of the population is in the
level $|{2} \rangle$ (continuous line). In both cases the Rabi frequency of the pump laser is 0.8
$\gamma_{3}$. } \label{fig:confronto}
\end{figure}

\begin{figure}[!]
\onefigure[width=.4\textwidth ]{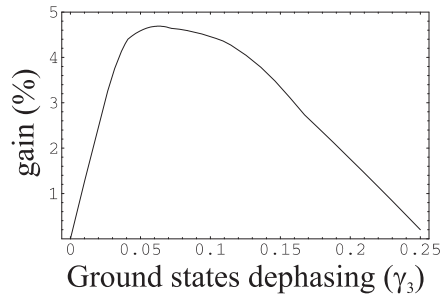} \caption{Percentage gain as a function of the
ground state dephasing $\gamma_1$ for the same experimental conditions as in figure
\ref{fig:confronto}.} \label{fig:gain}
\end{figure}

We can now check that the steady state atomic susceptibility computed from equation (\ref{eq:chi})
is the same as the one computed from the full solution of the density matrix for the parameters
considered here. This is shown in figure \ref{fig:confronto} left for a pump Rabi frequency of $0.8
\gamma_3$ and $30\%$ of the atomic population in level $|{2}\rangle$. This corresponds to the
situation where $\Gamma_1=0.2 \gamma_3$.
We denote by $\delta_{p} =\omega_{31}-\omega_{p}$ the probe detuning while the pump beam is taken
to be exactly at resonance ($\omega_{P}=\omega_{32}$).
From equations (\ref{eq:chi}) and (\ref{eq:T}) we obtain the transmission spectrum around the
atomic resonance of the probe laser through a 10 cm long cell containing $^{87}$Rb at a temperature
of $35^\circ$ C. This is reported in figure \ref{fig:confronto} right for a pump Rabi frequency of
$0.8 \ \gamma_3$. The transmission shows a peak in correspondence to the probe resonance, i.e. when
the pump and probe lasers close a Raman transition between levels $|{1} \rangle$ and $|{2}
\rangle$. This is precisely the electromagnetically induced transparency (EIT) in which the probe
absorption at resonance is canceled by destructive quantum interference between the two possible
absorption paths for the probe laser, namely the two--step transition from $|{1} \rangle$ to level
$|{2} \rangle$ through the excited level $|{3} \rangle$ and the Raman two--photon transition
between levels $|{1} \rangle$ and $|{2} \rangle$ \cite{pavone,ari}. The peak in the spectrum goes
all the way to full transmission when all the atomic population is placed in level $|{1} \rangle$
(dashed line). On the contrary, when some population ($30\%$ in figure \ref{fig:confronto} right)
is present in level $|{2} \rangle$ the transmission goes above unity indicating the presence of
gain (continuous line). We should remark that, as shown in figure \ref{fig:populations}, no
population inversion is present in the system therefore we are fulfilling the condition for gain
without inversion \cite{ari}. In the situation considered here increasing the population of level
$|{2} \rangle$ does not necessarily lead to extracting more gain. Indeed since we are changing the
population ratio by incoherently removing population from level $|{1} \rangle$, which in turn
increases the dephasing rate $\gamma_1$ between the ground sublevels. When the dephasing increases
the EIT effect is reduced and no gain is observed. In figure \ref{fig:gain} we report the
centerline gain as a function of the dephasing rate $\gamma_1$ with all the other experimental
parameters fixed as in figure \ref{fig:confronto}. As expected the gain increases to a maximum and
then drops down to zero.

\begin{figure}[!]
\onefigure[width=.5\textwidth ]{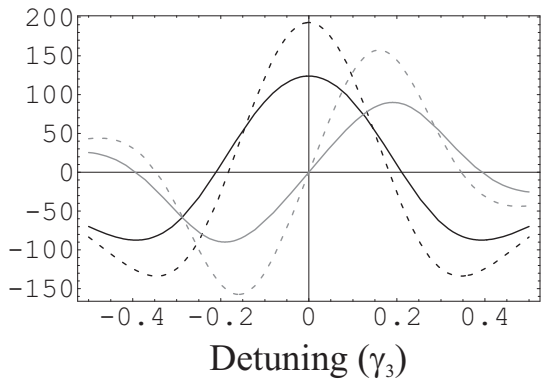} \caption{Reciprocal group velocity (black line) and
group velocity dispersion function $\cal{D}$ (gray line) [(m/s)$^{-1}$] as a function of probe
detuning $\delta_p$ (in unit of $\gamma_3$) for the same experimental conditions as in fig.
\ref{fig:confronto}. Dashed lines correspond to the case when all population is in level $|{1}
\rangle$, continuous lines correspond to the case when 30\% of the population is in level $|{2}
\rangle$. Gray curves are reduced by a factor of 10.} \label{fig:group}
\end{figure}

Combining equations (\ref{eq:vg}) e (\ref{eq:dg}) with equation
(\ref{eq:chi}), we obtain the frequency dependence of the group
velocity and its dispersion around the probe resonance for the
same experimental parameters as before as reported in figure
\ref{fig:group}. We note that there are three probe frequencies
where the group velocity dispersion vanishes. These points
correspond to frequency values where a suitable probe pulse can
propagate through the medium without distortion. One of these
points corresponds to the line center where we have retarded
propagation. The other two, which are symmetrical with respect to
the first point, correspond to anomalous propagation, i.e. a
negative group velocity. We note that, when some population is
placed in level $|{2} \rangle$, the positive minimum of the group
velocity is increased. This is not an effect of population but of
the increase in dephasing of the ground sub--levels. At the same
time the negative minimum group velocity is also increased. This
means that, under the conditions of gain without inversion, a
propagating pulse will be slowed down while, at the same time,
undergoing amplification whereas in the anomalous propagation
region the pulse is advanced but not amplified. However both the
pulse delay and the pulse advance are reduced with respect to
normal EIT.

\begin{figure}[!]
\includegraphics[width=.51\textwidth ]{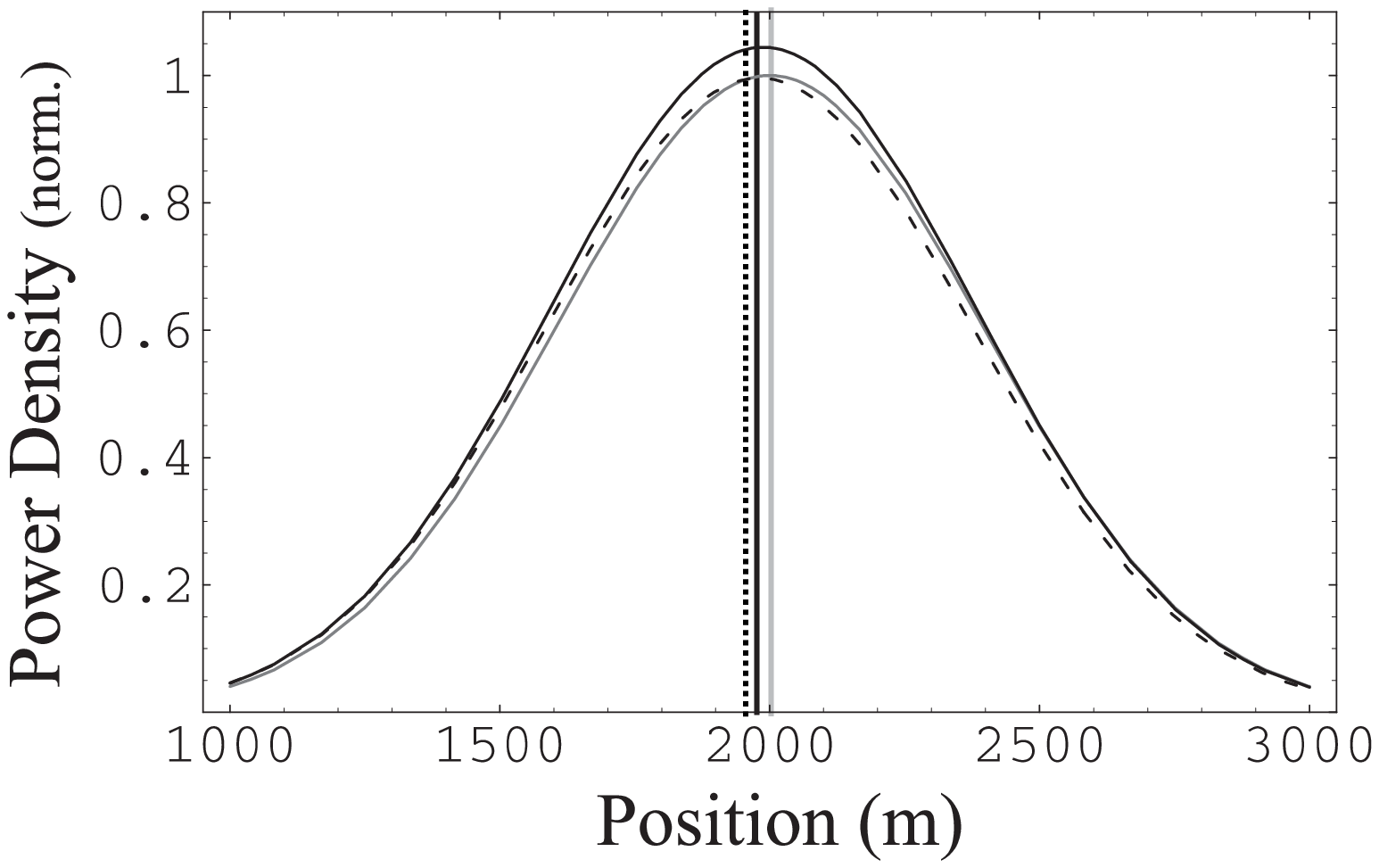}
\includegraphics[width=.48\textwidth ]{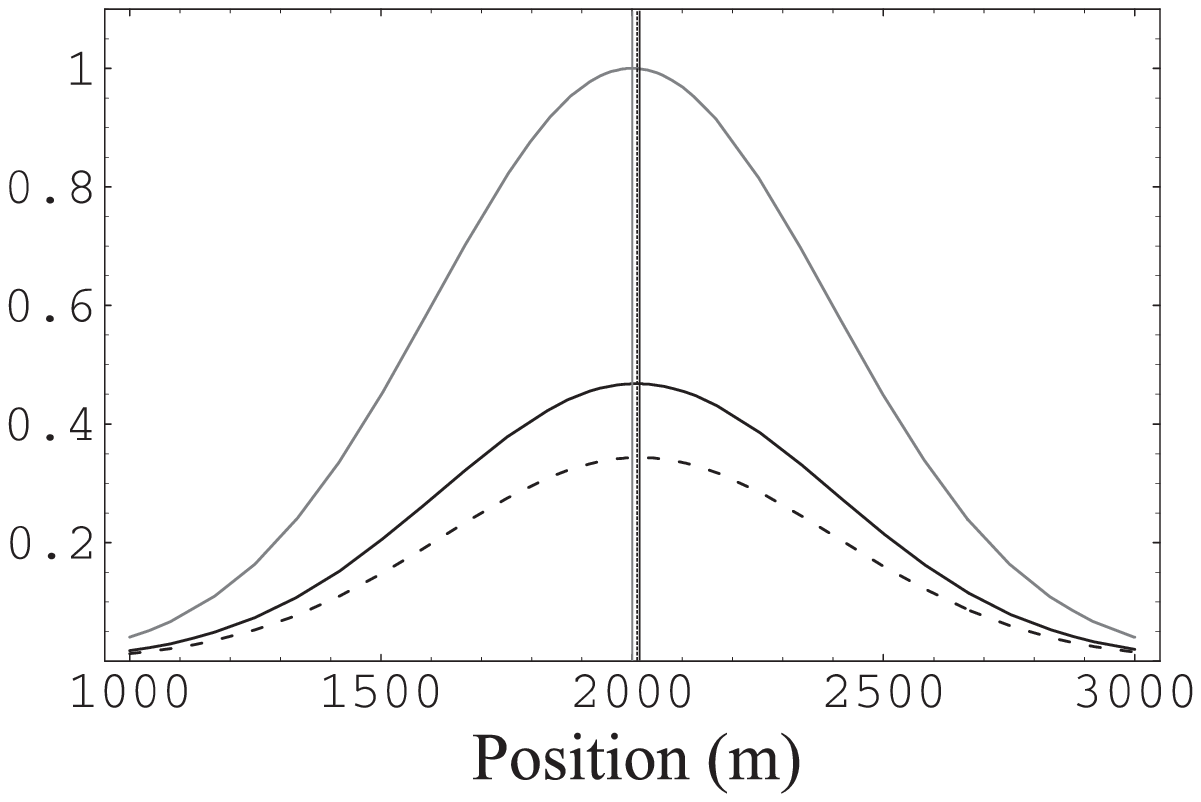}
\caption{Power density for a Gaussian probe pulse normalized to the pulse propagating in vacuum. On
the left, the peak central frequency is on resonance resulting in retarded pulse propagation. The
vertical lines in the plot indicate the center of masses of the pulses. On the right, the probe
detuning corresponds to the negative minima of reciprocal group velocity showing advanced
propagation. The gray line refers to the pulse propagating in vacuum, the dashed line to the case
where all population is in the level $|{1} \rangle$ and the black continuous line to the case where
$30\%$ of the population is in the level $|{2} \rangle$. Experimental parameters are the same as
the fig. \ref{fig:confronto}.} \label{fig:propag}
\end{figure}

\begin{figure}[!]
\includegraphics[width=.505\textwidth ]{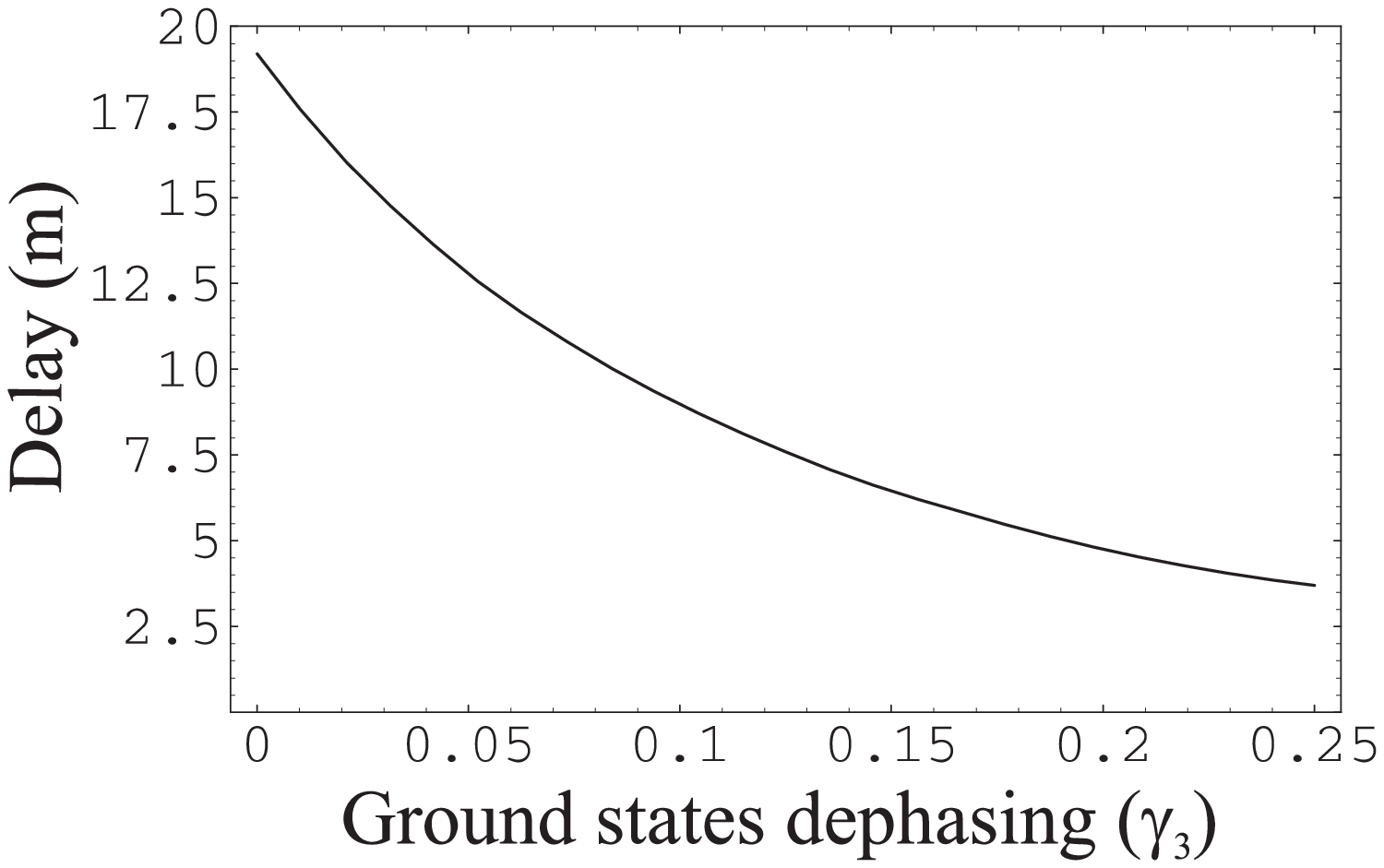}
\includegraphics[width=.485\textwidth ]{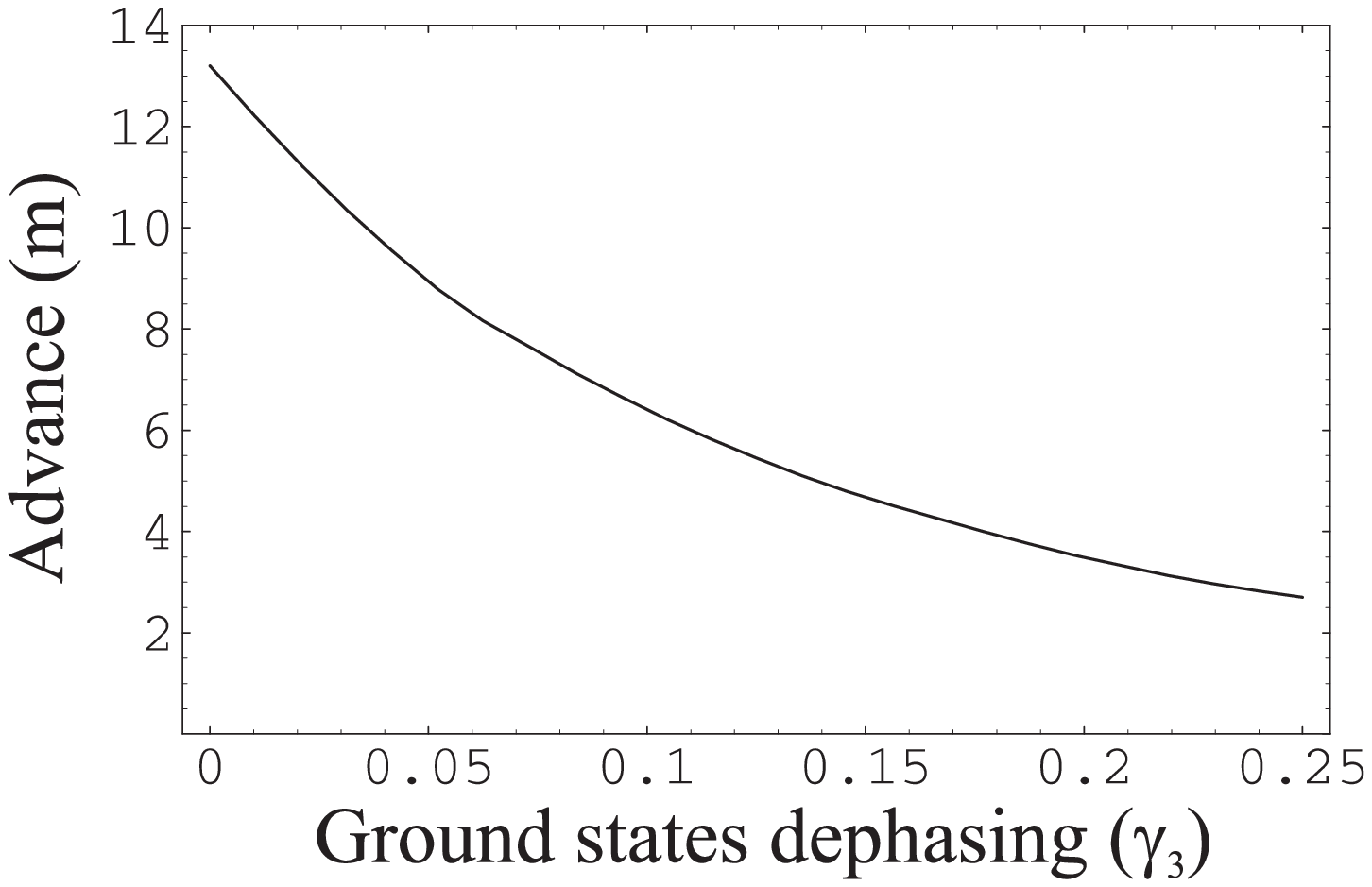}
\caption{On the left pulse delay, on the right pulse advance as a
function of the dephasing rate $\gamma_1$ for the same
experimental parameters as in figure \ref{fig:propag}}
\label{fig:rmax}
\end{figure}

In figure \ref{fig:propag}, we report the power density spectrum for a Gaussian probe pulse
propagating through a 10 cm long cell again for a pump Rabi frequency of $0.8 \ \gamma_3$. On the
left we show the resonant case where the pulse propagation is retarded. We have chosen the
parameters to be at the maximum amplification (black line), in such a case the delay with respect
to a pulse propagating in vacuum (grey line) is 12.2 m which amounts to a velocity of
$\frac{c}{120}$ in the cell. In the normal EIT situation (dotted line) with all the population in
level $|{1} \rangle$ and virtually no dephasing between the ground sublevels this delay is 18.9 m.
As already mentioned, when we increase the decay rate from level $|{1} \rangle$, there is a
reduction of the delay as a consequence of the larger dephasing between the ground levels. As
reported in figure \ref{fig:rmax} (left) for our experimental parameters the delay is reduced below
5 m when the dephasing is equal to 0.25 $\gamma_3$. As reported in figure \ref{fig:gain} at the
same level of dephasing no amplification is observable.

Conversely the detuned case, shown in figure \ref{fig:propag}
right, exhibits a pulse advanced of 13.3 m with respect to the one
propagating in vacuum, but the advanced propagation is accompanied
by absorption. The absorption is reduced when some population is
placed in level $|2 \rangle$ but, at the same time, also the
advance is reduced to 8.6 m. We note that the observed delay and
advance can be greatly enhanced by reducing the pump Rabi
frequency $\Omega_P$ but the effect of the levels population
balance on the pulse advance is strongly suppressed. In the
anomalous propagation regime, the pulse advance is decreased with
respect to the normal EIT in a such way that the pulse edge never
appears ahead of the edge of a pulse propagating in vacuum for the
same distance. This effect is enhanced by the larger dephasing
rate as shown in figure \ref{fig:rmax} (right) and the advance is
below 3 m when the dephasing rate reaches 0.25 $\Gamma_1$. We note
that in the regime of population inversion the amplified pulse
leading edge can indeed precede that of the vacuum propagating
pulse. This is not surprising since, by inverting the atomic
levels population, we are effectively storing energy in the medium
\cite{artoniPRA}.

In conclusion we have shown that in the presence of gain without
inversion it is possible to observe effects similar to those of
pure EIT, i.e. retarded propagation in absence of absorption for
the resonant case and anomalous propagation accompanied by
absorption in the detuned case. However in presence of gain
without inversion a propagating pulse is less retarded than in the
pure EIT case, while at the same time undergoing amplification. On
the other hand in the anomalous propagation region the pulse
absorption is reduced and the advance is also reduced in such a
way that the pulse edge never precedes that of a pulse propagating
in vacuum. We have performed all the calculations for a sample of
a $^{87}$Rb vapor at $35^\circ$ C in a 10 cm long cell. We have
examined a realistic model to create the atomic population ratios
needed for AWI introducing an incoherent loss rate from one of the
ground levels. We have discussed how this mechanism affects the
dephasing rates of the ground levels. This model is easily
extendable to the case of weak coherent fields to study
decoherence in quantum memories as well as to discuss AWI in
connection with photon cloning.

\acknowledgments We thank M. Inguscio and E. Rimini for most valuable discussions and their
continuous support of this work. We are deeply indebted to M. Artoni and G. La Rocca for their many
useful suggestions.

\end{document}